\documentclass[pra,print,superscriptaddress,twocolumn]{revtex4}
\usepackage{amssymb}
\usepackage{amsmath}
\usepackage{epsfig}
\usepackage{color}
\usepackage{graphics, graphicx}
\usepackage{bbold}
\usepackage{psfrag}
\usepackage{mathcomp}
\usepackage{subfigure}
\usepackage{verbatim}
\usepackage{color}
\usepackage[colorlinks,citecolor=blue]{hyperref}

\begin{document}

\title{Bosonic Peierls state emerging from the one-dimensional
Ising-Kondo interaction}
\author{Jingtao Fan}
\affiliation{State Key Laboratory of Quantum Optics and Quantum Optics Devices, Institute
of Laser Spectroscopy, Shanxi University, Taiyuan 030006, China}
\affiliation{Collaborative Innovation Center of Extreme Optics, Shanxi University,
Taiyuan 030006, China}
\author{Xiaofan Zhou}
\thanks{zhouxiaofan@sxu.edu.cn}
\affiliation{State Key Laboratory of Quantum Optics and Quantum Optics Devices, Institute
of Laser Spectroscopy, Shanxi University, Taiyuan 030006, China}
\affiliation{Collaborative Innovation Center of Extreme Optics, Shanxi University,
Taiyuan 030006, China}
\author{Suotang Jia}
\affiliation{State Key Laboratory of Quantum Optics and Quantum Optics Devices, Institute
of Laser Spectroscopy, Shanxi University, Taiyuan 030006, China}
\affiliation{Collaborative Innovation Center of Extreme Optics, Shanxi University,
Taiyuan 030006, China}

\begin{abstract}
As an important effect induced by the particle-lattice interaction, the
Peierls transition, a hot topic in condensed matter physics, is usually
believed to occur in the one-dimensional fermionic systems. We here study a
bosonic version of the one-dimensional Ising-Kondo lattice model, which
describes itinerant bosons interact with the localized magnetic moments via
only longitudinal Kondo exchange.\ We show that, by means of perturbation
analysis and numerical density-matrix renormalization group method, a
bosonic analog of the Peierls state can occur in proper parameters regimes.
The Peierls state here is characterized by the formation of a long-range
spin-density-wave order, the periodicity of which is set by the density of the itinerant bosons. The ground-state phase diagram is mapped out by
extrapolating the finite-size results to thermodynamic limit. Apart from the
bosonic Peierls state, we also reveal the presence of some other magnetic
orders, including a paramagnetic phase and a ferromagnetic phase. We finally
propose a possible experimental scheme with ultracold atoms in optical
lattices. Our results broaden the frontiers of the current understanding of
the one-dimensional particle-lattice interaction system.
\end{abstract}

\maketitle

\section{Introduction}

The interaction between particle and lattice degrees of freedom is a
fundamental aspect of condensed matter physics, with far-reaching
implications for our understanding and engineering of materials. At the
heart of this interplay lies the complex dance between the microscopic
constituents of solids - the particles, such as electrons and exciton, and
the periodic arrangement of atoms known as the lattice~\cite{LP1}. This
particle-lattice interaction is the key to unlocking the rich tapestry of
various quantum phenomena, from the emergence of electronic band structures
to the intricate mechanisms underlying superconductivity and phase
transitions~\cite{LP2,LP3,LP4,LP5}. Especially in one-dimensional metals, the
electron-lattice interaction can destabilize the original Fermi surface,
leading to a lattice distortion in the form of density waves, known as the
Peierls distortion~\cite{Peierls1}. While the mechanism underlying the
Peierls transition in solids depends on the Fermi statistics of particles,
it has been shown recently that similar effects can also be presented in
some bosonic systems with tailored boson-lattice interaction~\cite%
{BP1,BP2,BP3,BP4}. Given that the Peierls transition in fermionic systems
has been extensively investigated, its understanding in the context of
bosonic systems is far from complete. Moreover, the models proposed in Refs.~\cite{BP1,BP2,BP3,BP4} are somehow artificially designed, making them less
inspired in bridging analogous phenomena observed in solids. Therefore, it
is highly desired to know weather the Peierls state can arise in any bosonic
analogs of existing solid-state systems which are relevant to real materials.

The interaction between mobile particles and fixed magnetic impurities,
known as Kondo physics, is of central importance in condensed matter physics%
~\cite{KLM1}. The Kondo interaction is responsible for a variety of
interesting phenomena, including the heavy fermion state~\cite%
{HeavyF1,HeavyF2}, the non-fermi liquid behavior~\cite%
{NonFermi1,NonFermi2,NonFermi3}, and the magnetic anisotropy effect~\cite%
{MagAni1,MagAni2,MagAni3}. Things become more interesting if the magnetic
centers are periodically distributed on a lattice and only longitudinal
Kondo exchange is considered, a situation where the Ising-Kondo lattice
model (IKL) works~\cite%
{IKL01,IKL1,IKL02,IKL03,IKL04,IKL2,IKL3,IKL4,IKL5,IKL6,IKL7,IKL8}. Since it
was first proposed to describe the concurrence of large specific heat jump
and weak antiferromagnetism in URu$_{2}$Si$_{2}$~\cite{IKL01}, the IKL has
been an important topic in condensed matter physics thanks to its simple
Ising-type coupling form, amenable to both analytical and numerical
treatments~\cite{IKL5}, and to its relevance to a series of real materials%
~\cite{IKL2,IKL3,IKL4}. Note that the localized magnetic moments in the IKL
are encoded in the lattice sites with spin fluctuations, behaving much like
phonons in a metal. Hence, the IKL is emerging as a system describing
itinerant particles moving on a dynamical lattice with the particle-lattice
interaction controlled by the Kondo coupling. The study of particle-lattice
problem therefore becomes very relevant in this context.

Armed with the ability to precisely control the inter-atomic interactions,
external potential fields, and artificial magnetic fields, ultracold atoms
in optical lattices have provided an important experimental platform for
exploring strongly correlated quantum many-body systems~\cite%
{AtomOL1,AtomOL2,AtomOL3,AtomOL4,AtomOL5}. Especially by loading atoms on
different bands of the optical lattices, it is possible to build systems
composed of both localized and mobile particles, analogs of Kondo physics
but with the possibility to use bosonic particles instead of fermionic ones%
~\cite{HighOL1,HighOL2,HighOL3,HighOL4,HighOL5,HighOL6,HighOL7,HighOL8}. With
these rapid experimental advances, it is interesting to consider a Kondo
interaction in which the electrons are replaced by spin-$1/2$ bosons~\cite%
{BHighOL1,BHighOL2,BHighOL3}. Hopefully, considering the analogy between the
IKL and the particle-lattice system, the change of particle statistics may
have potential implications for the existing Peierls theory.

In this paper, we study the 1D IKL with interacting spin-$1/2$ bosons
instead of fermions. The interplay between different kinds of interactions
reveals a variety of magnetic orders and nontrivial transitions between
them. Apart from the uniform paramagnetic phase (PM) and ferromagnetic phase (FM), the most interesting finding is the presence of a bosonic analog of
the Peierls state, which breaks the translational symmetry of the underlying
lattice. The bosonic Peierls state is characterized by the formation of a
long-range spin-density-wave (SDW) order, the period of which is uniquely
determined by the bosonic density. Adopting the numerical density-matrix
renormalization group (DMRG) method~\cite{dmrg1,dmrg2}, we map out the ground-state phase
diagrams in various parameter spaces. The bosonic Peierls state is shown to
appear at intermediate values of the Kondo coupling, implying that it is
basically a non-perturbative effect and depends on nontrivial competitions
between different energy scales. We also propose a possible experimental
scheme with ultracold atoms in optical lattices.

This work is organized as follows. In Sec. II, we describe
the proposed model and present the Hamiltonian. In
Sec. III, we provide a perturbative analysis, accurate in two opposite limits, for the bosonic IKL.
In Sec. IV, we numerically calculate relevant correlation functions to specify different long-range orders, and map out the ground-state phase diagrams in various parameter spaces. We discuss the possible experimental implementations of our model in Sec. V and summarize in
Sec. VI.

\section{Model and Hamiltonian}

\label{sec:system}

The model we consider is a bosonic version of the 1D IKL, which was
originally proposed in the context of heavy-fermion compounds~\cite{IKL01}.
Similar to the fermionic system, the bosonic IKL considered here concerns
conduction bosons interacting with localized magnetic moments via Ising-type
(longitudinal) Kondo exchange. The Hamiltonian describing such a system can
be written as ($\hbar =1$ throughout)%
\begin{eqnarray}
H &=&-t\sum_{\left\langle i,j\right\rangle ,\sigma }\hat{b}_{i,\sigma
}^{\dag }\hat{b}_{j,\sigma }+\frac{U}{2}\sum_{j,\sigma }\hat{n}_{j,\sigma }(%
\hat{n}_{j,\sigma }-1)  \notag \\
&&+\mu \sum_{j,\sigma }\hat{n}_{j,\sigma }+J\sum_{j}\hat{s}_{j}^{z}\hat{S}%
_{j}^{z}+h\sum_{j}\hat{S}_{j}^{x}  \label{H1}
\end{eqnarray}%
where $\hat{b}_{j,\sigma }^{\dag }$ ($\hat{b}_{j,\sigma }$) is the creation
(annihilation) field operator of the conduction boson with spin $\sigma $ (=$%
\uparrow ,\downarrow $) at lattice site $j$, and $\hat{n}_{j,\sigma } \!\!=\!\! %
\hat{b}_{j,\sigma }^{\dag }\hat{b}_{j,\sigma }$ is the corresponding boson
number operator. $\hat{S}_{j}^{z}$ and $\hat{S}_{j}^{x}$ are the spin-1/2
operators for the localized magnetic moment. The spin operator $\hat{s}%
_{j}^{z}$\ for the conduction bosons is defined by $\hat{s}%
_{j}^{z}=\sum_{\tau ,\tau ^{\prime }}^{\dag }\hat{b}_{j,\tau }^{\dag }\sigma
_{\tau ,\tau ^{\prime }}^{z}\hat{b}_{j,\tau ^{\prime }}/2$ where $\sigma ^{z}
$ is the Pauli-$z$ matrix. With these definitions, the first three terms of
Eq. (\ref{H1}) constitute the conventional Bose-Hubbard Hamiltonian with the
particle hopping rate $t$ between adjacent sites $\left\langle i,j\right\rangle$, the repulsive Hubbard interaction $U$, and the
chemical potential $\mu $. The fourth term of Eq. (\ref{H1}) is the
so-called Kondo coupling term which describes the Ising-type interaction
between conduction bosons and localized magnetic moments. The coupling
constant $J$ is assumed to be positive, implying an antiferromagnetic Ising
interaction. The ferromagnetic coupling with $J<0$ is physically equivalent
to the antiferromagnetic case in the sense that they are connected by a spin
rotation of $\pi $ along the spin-$x$ axis. Note that a transverse field $h$
introduces the dynamics of the localized magnetic moments through the last
term of Eq. (\ref{H1}). In the following discussion, we set the energy scale
by taking $t=1$, and we also take $h=1$ unless otherwise specified.

The Hamiltonian (\ref{H1}) can be alternatively viewed as a minimal model
describing the strongly correlated spinful bosons moving on a dynamical
lattice~\cite{BP1,Larson20}. Here, the lattice degree of freedom is
dynamically generated by spin excitations of the localized magnetic moments
instead of phonons in traditional condensed matter materials. In this
context, the transverse field $h $ plays the role of phonon energies,
which makes the lattice nonadiabatic. For $J\ll 1$, the bosons are nearly
decoupled from the localized moments and the system behave much like that of
the Bose-Hubbard model, which supports a Mott insulator (MI) and a
superfluid phase, both of which are uniform in space. At relatively large Kondo coupling, however, the mobile
bosons are deeply dressed by the lattice degree of freedom and some other
self-organized orders may develop. As a matter of fact, for the 1D fermionic
IKL with a two-point Fermi surface, a density-wave instability can occur at strong Kondo coupling~\cite{IKL8}. This
effect turns out to be a consequence of both the perfect Fermi surface
nesting and the\ Fermion-lattice interaction. It is well known that the
Fermi surface structure is absent for the bosonic model considered here, one can thus naively expect that the density-wave order should consequently vanish for the Hamiltonian (\ref{H1}).  However, as will be
elucidated in the subsequent sections, the Kondo coupling and the Hubbard interaction between conduction bosons may cooperate in some nontrivial way yielding analogous orders that break the translational symmetry of the underlying lattice.

Here, we perform the state-of-the-art DMRG calculations to calculate the many-body ground-states of the system under open boundary conditions.
In our numerical simulations, the density $\rho =N/L$ is a
good quantum number which can be varied from zero to two. Here $N$ is the total number of conduction bosons, and $L$ is the lattice size.
We set the cutoff of single-site atom number as $n_\text{cutoff}=4$. The effect of such a cutoff can be neglected for strong interaction, while in the weak interaction region, the phases and phase boundaries may be slightly affected by the cutoff. We find that $n_\text{cutoff}=4$ is enough to determine the phase boundaries.
We set lattice size up to $L=32$, for which we retain 800 truncated states per DMRG block and perform 40 sweeps with a maximum truncation error $\sim10^{-9}$.

\section{Perturbation theory}

\label{sec:perturbation theory}

Before providing the numerical results for general parameters, it is
beneficial to put the interaction constant $J$ and $U$ to two opposite
limits, $J,U\ll 1$ and $J,U\gg 1$, in which the perturbation treatment
becomes accurate. In the following, we derive effective spin Hamiltonians in
the two limits, respectively. This effective description not only provide a
clear physical picture to understand mechanisms behind the formation of
various symmetry-broken phases, but also reinforces and complements the
numerical results that will be subsequently presented.

\subsection{weak coupling}

\label{sec:weak}

We first focus on the weak coupling case where $J$ and $U$ can be treated as
perturbation compared with the hopping rate $t$. When $J,U=0$, the conduction
bosons are decoupled from the local moments and can freely move on the
lattice with dispersion relation $\epsilon _{k}=-t\cos ka$. The full Hilbert
space of the conduction bosons is thus conveniently spanned by various
productions of the single-particle eigenstates $\psi _{k}(j)$. The ground
state is a condensate formed by putting all the bosons onto the bottom of
the energy band. Due to the free choice of the spin up and down of the bose
condensate, the ground state has a degeneracy of $N+1$. This spin degeneracy
can be lifted by the Kondo coupling at the first order of $J/t$. This can be
clearly seen by projecting the Hamiltonian onto the degenerate ground-state
subspace, which yields%
\begin{equation}
H^{(1)}=J\hat{s}^{z,0}\sum_{j}\hat{S}_{j}^{z}+h \sum_{j}\hat{S}_{j}^{x}
\label{Hw1}
\end{equation}%
where
\begin{equation}
\hat{s}^{z,0}=\frac{1}{2}\sum_{i,l}\sum_{\tau ,\tau ^{\prime }}\Xi
_{0}^{0}\Xi _{l}^{0}\hat{c}_{i,\tau }^{\dag }\sigma _{\tau ,\tau ^{\prime
}}^{z}\hat{c}_{i+l,\tau ^{\prime }}/2  \label{sz}
\end{equation}%
is the spin operator for a conduction boson in its single-particle ground
state, and we have defined the correlation function $\Xi _{l}^{k}=\psi
_{k}(0)\psi _{k}^{\ast }(l)$. Note that in deriving Eqs. (\ref{Hw1}) and (\ref%
{sz}), we have used the translational invariance of the system, and the
contribution of the Hubbard interaction $U$ is dropped since it does not
lift the spin degeneracy at this order. The first term of the effective
Hamiltonian (\ref{Hw1}) describes a collective Ising-type interaction, which
ferrromagnetically aligns the condensate spins. These aligned condensate
spins in turn antiferromagnetically couple to the total spin of the local
moments. The transverse field $h$, on the other hand, induces quantum
fluctuations along the ordered axis, which, if dominates, may break the ferromagnetism built up by the Kondo coupling.

The virtual excitations above the Bose condensate may contribute to the
energy through some higher order processes. Employing the standard
perturbation theory, the correction to the effective Hamiltonian at the
second order of the Kondo exchange reads~\cite{BHighOL2}

\begin{equation}
H^{(2)}=-NJ^{2}\sum_{j,l}R_{l}\hat{S}_{j}^{z}\hat{S}_{j+l}^{z}+2J^{2}R_{0}%
\hat{s}^{z,0}\sum_{j}\hat{S}_{j}^{z}  \label{Hw2}
\end{equation}%
where $R_{l}=\sum_{k}\Xi _{l}^{k}\Xi _{l}^{0\ast }/(\epsilon _{k}-\epsilon
_{0})$. While the second term of the Hamiltonian (\ref{Hw2}) simply
generates a correction of the coupling constant $J$ in $H^{(1)}$, the first
term is the bosonic analog of the Ruderman-Kittel-Kasaya-Yosida (RKKY) interaction which, in solids, is mediated
by the conduction electrons. In the fermionic Kondo lattice model, the RKKY interaction
oscillates around zero at a characteristic length which is inversely
proportional to the Fermi momentum $k_{F}$. Therefore, an oscillatory magnetic
ordering naturally emerges in both the conduction electrons and the local
moments there. However, such length scale disappears for the bosonic model
here due to the absence of the Fermi surface. In fact, it can be shown that
the coupling function $R_{l}$ is strictly positive for any $l$, favoring
parallel alignment of the localized spins (the spins of the conduction bosons align
opposite to the localized spins). It is important to notice that, for 1D,
the coupling constant $NR_{l}$ diverges with the system size as $l$
approaches zero.\ Although this divergence actually invalidates the
perturbation expansion in terms of $J$, it still suggests that the
ferromagnetic ordering of both the localized and the itinerant spins
can be formed within themselves at weak coupling. This picture has also been confirmed by numerical calculations.

\subsection{strong coupling}

\label{sec:strong}

We turn to the opposite limit where $J,U\gg 1$, and focus on the commensurate filling at which $\rho$ takes integer values. The ground state in this regime is a MI.
At $t=h=0$,\ the MI can be written as the following product state
\begin{equation}
\left\vert \text{MI}_{\rho ,m}\right\rangle = \prod \limits_{j}\otimes
\left\vert \frac{\rho }{2}+\bar{\sigma}_{j}m,\frac{\rho }{2}-\bar{\sigma}%
_{j}m\right\rangle _{b}\left\vert -\bar{\sigma}_{j}\right\rangle _{s}.
\label{MI}
\end{equation}%
In Eq. (\ref{MI}), the subscripts $b$ and $s$ represent conduction bosons
and localized spins, respectively. $m$ ($\geqslant 0$) denotes the magnetization of the
conduction bosons per site along spin-$z$ component, and the occupation of
spin-up (spin-down) bosons at site $j$ is thus $\rho /2+\bar{\sigma}_{j}m$ ($%
\rho /2-\bar{\sigma}_{j}m$) with $\bar{\sigma}_{j}=\pm 1$ specifying the
spin orientation. The minus sign in front of $\bar{\sigma}_{j}$ in the
localized spin state manifests the antiferromagnetic nature of the Kondo
coupling. From above definition, the maximum value that $m$ can take is $%
\rho /2$. The energy per site of $\left\vert \text{MI}_{\rho
,m}\right\rangle $ is $E_{\rho ,m}=U(\rho ^{2}/4-\rho /2+m^{2})-mJ/2$, from
which we see that increasing $m$ may lower the Kondo energy but elevate the
Hubbard interaction. Minimizing the total energy with respect to $m$ yields
\begin{equation}
m=\left\{
\begin{array}{ll}
\min [\frac{\rho }{2},\mathcal{F(}\frac{J}{4U}+\frac{1}{2}\mathcal{)}]\text{%
, } \text{for\ }\rho /2\in \text{integer} \\
\min [\frac{\rho }{2},\mathcal{F(}\frac{J}{4U}\mathcal{)}+\frac{1}{2}]\text{%
, } \text{for\ }\rho /2\in \text{half integer}%
\end{array}%
\right.   \label{Mz}
\end{equation}%
where the function $\mathcal{F(}x\mathcal{)}$ extracts the integer part of $x
$. To determine the spectral properties of the system one can calculate the
so-called charge gap $\Delta $ defined as
\begin{equation}
\Delta =E_{0}(N+1,L)+E_{0}(N-1,L)-2E_{0}(N,L)
\end{equation}%
where $E_{0}(N,L)$ is the ground-state energy for a system of length $L$
with particle number $N$. With the above results, it is straightforward to
deduce the charge gap $\Delta $,
\begin{equation}
\Delta =\left\{
\begin{array}{ll}
\! U(1-2m)+\frac{J}{2}, \text{ for }J/4U<m \\
\! U(1+2m)-\frac{J}{2}, \text{ for }J/4U\geqslant m\text{ and }m<\rho /2 \\
\! U\text{,} \text{ for }J/4U\geqslant m\text{ and }m=\rho /2.%
\end{array}%
\right.
\end{equation}%
It follows that the behavior of $\Delta $ relies on the parity of the density $\rho $: as $J$ $\longrightarrow 0$, the charge gap vanishes
for odd $\rho $ whereas it approaches $U$ for even $\rho $. It is then clear
that for even $\rho $, the limit $U\gg 1$ alone can guarantee the MI nature
of the ground state, but for odd $\rho $ the existence of MI requires both $%
U\ $and $J$ are sufficiently large compared to $t$.

There is a remaining degeneracy for the MI state (\ref{MI}), associated with
the spin orientation $\bar{\sigma}_{j}$ at each lattice site. This
degeneracy can be removed by taking into consideration the hopping process
of the conduction bosons through some virtual excited states. Adiabatically
eliminating these virtual excitations in the second order with respect to $%
t/J$ and $t/U$, we obtain an effective spin Hamiltonian (see Appendix for details on the derivation)
\begin{eqnarray}
H &=&\mathcal{J}\sum_{<i,j>}\hat{s}_{i}^{z}\hat{s}_{j}^{z}+4m^{2}\mathcal{J}%
\sum_{<i,j>}\hat{S}_{i}^{z}\hat{S}_{j}^{z}  \notag \\
&&-2m\mathcal{J}\sum_{<i,j>}\hat{s}_{i}^{z}\hat{S}_{j}^{z}+\mathcal{J}%
^{\prime }\sum_{j}\hat{s}_{j}^{z}\hat{S}_{j}^{z}  \label{Hs}
\end{eqnarray}%
where
\begin{eqnarray}
\mathcal{J} &=&\frac{1}{4m^{2}}\left[ \frac{2t^{2}}{J+(2-4m)U}\right.
\label{J} \\
&&\left. +\frac{t^{2}}{(2m+1)U}-\frac{2t^{2}}{U}\right] ,  \notag
\end{eqnarray}%
and%
\begin{eqnarray}
\mathcal{J}^{\prime } &=&\frac{1}{2m}\left[ \frac{4t^{2}}{J+(2-4m)U}\right.
\label{Jp} \\
&&\left. +\frac{2t^{2}}{(2m+1)U}+\frac{4t^{2}}{U}+\frac{h^{2}}{mJ}\right]
\notag
\end{eqnarray}

With the effective Hamiltonian (\ref{Hs}), it is much easier to reveal the
ground-state magnetism in different parameter regimes. While the
first (second) term of the Hamiltonian (\ref{Hs}) describes the direct Ising
interactions within conduction bosons (localized spins), the last two terms
account for cross couplings between conduction bosons and localized spins,
which are respectively responsible for the neighboring and on-site spin
interactions. It can be shown that $\mathcal{J}^{\prime }\ $is strictly
positive, which is understood since the antiferromagnetic nature of the
on-site Kondo coupling should always be preserved. The sign of the coupling
constant $\mathcal{J}$, however, depends on the system parameters and
ultimately determines the magnetic orders for both conduction bosons and
localized spins. By requiring $\mathcal{J}>0$, we obtain the condition where
the antiferromagnetic phase (AFM) is formed, i.e.,
\begin{equation}
4m-2<\frac{J}{U}<\frac{16m^{2}}{4m+1},\text{ (}m\neq 0\text{)}.  \label{AF}
\end{equation}%
In this phase, the spin orientation alternates in the real space, which corresponds to an ordering wave vector $k^{\max }=\pi $. The stable condition for the FM can
be similarly derived by setting $\mathcal{J}<0$, and we have%
\begin{equation}
\frac{J}{U}>\frac{16m^{2}}{4m+1},\text{ (}m\neq 0\text{)}.  \label{FM}
\end{equation}%
It follows from Eq. (\ref{Mz}) that at even $\rho $, the magnetization $m$
vanishes for $J<2U$, giving rise to the formation of PM. Since
the value of $m$\ depends on $\rho $, the coupling constant $%
\mathcal{J}$, $\mathcal{J}^{\prime }$ and the related equations (\ref{AF})-(%
\ref{FM}) are basically density dependent. As an example, in the case of unit
filling with $\rho=1 $, we have $m\equiv 1/2$. Consequently, the conduction bosons
construct AFM for $J/U<4/3$ and FM for $J/U>4/3$.

\section{Long-range order}

We are now at the stage to explore the ground-state properties under general
parameter regimes, which we resort to the numerical calculations based on the
DMRG algorithm. To quantitatively characterize different magnetic orders and the transitions between them, we consider the spin structure factor, defined by

\begin{equation}
S(k)=\frac{1}{L}\sum_{l,j}\left\langle \hat{s}_{l}^{z}\hat{s}%
_{j}^{z}\right\rangle e^{i(l-j)k}.
\end{equation}%
The scaled spin structure factor $\lim_{L\rightarrow \infty }$ $S(k)/L$
develops a peak at some ordering wave vector $k^{\max }$ in the presence of
long-range order~\cite{Long}. While $k^{\max }$ determines the wave length of
a SDW through $\lambda =2\pi /k^{\max }$, the height of the peak, defined by
$S^{\max }=\lim_{L\rightarrow \infty }$ $S(k^{\max })/L$, measures the
corresponding ordering strength. In this way, a $k^{\max }$-ordered SDW phase (FM)
is characterized by $S^{\max }\neq 0$ and $k^{\max }\neq 0$ ($S^{\max }\neq 0
$ and $k^{\max }=0$). If $S^{\max }$ vanishes in the thermodynamic limit,
the ground state owns no magnetic long-range order and is thus identified as
PM. Notice that owing to the space-inversion symmetry of the system, any
peaks in $S(k)$, if exist, should be exactly symmetric about $k=0$.
\begin{figure}[tp]
\includegraphics[width=9.0cm]{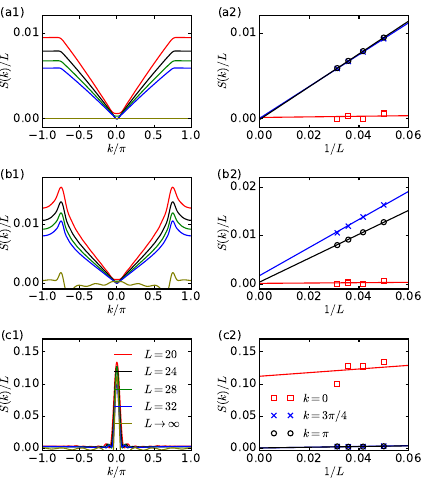}
\caption{(a1)-(c1) The scaled spin structure factor $S(k)/L$ and (a2)-(c2)
the corresponding finite-size scalings at three characteristic wave vectors
for systems with (a) $J=0.2$, (b) $J=10.0$, and (c)\ $J=25.0$. Different
system sizes are characterized by lines with different colors. In these
figures, we set $\protect\rho =0.75$ and $U=50$. }
\label{fig1}
\end{figure}

The left column of Fig.~\ref{fig1} plot $S(k)/L$ for different $J$ with $U=50
$ and $\rho =0.75$, and the results for different system sizes are labeled
by lines of different colors. The values of $L\rightarrow \infty $ are
obtained by the standard finite size scaling. The corresponding scaling
details for three representative wave vectors, $k/\pi =0,$ $ 0.75$, $ 1$, are
shown in the right column of Fig.~\ref{fig1}. We first pay attention to the
two limits, $J\ll 1 $ and $J\gg 1 $. As shown in Fig.~\ref%
{fig1}(a1), for $J=0.2$, the scaled structure factor generally decreases
with the system size and eventually vanishes for all the wave vectors,
which is the character of PM. The opposite limit with $J=25.0$ is
demonstrated in Fig \ref{fig1}(c1). In this case, the peak at $k/\pi =0$
overwhelms values at other wave vectors, indicating the formation of FM.
Something interesting happens if the coupling strength is tuned to an
intermediate value  $J=10.0$. As shown in Fig.~\ref{fig1}(b1),  a sharp peak of  $S(k)/L$ appears at $k/\pi =k^{\max }/\pi =\pm 0.75$, which
does not vanish even in the thermodynamic limit $L\rightarrow \infty $. This
is a clear signal of a SDW phase with the ordering wave vector $k^{\max
}=\pm 0.75\pi $. With these results, we can expect that by varying the coupling strength $J$ from
small to large, the system may undergo a transition from PM to SDW, and
then end up with FM.

\begin{figure}[b]
\includegraphics[width=9.0cm]{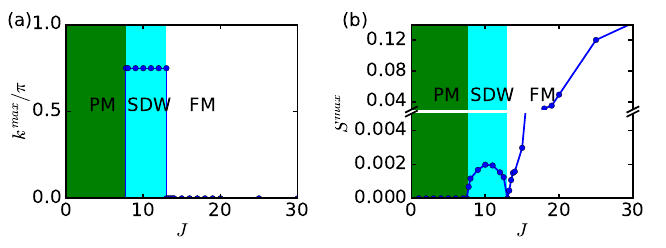}
\caption{(a) The ordering wave vector $k^{\max }$ and (b) the ordering
strength $S^{\max }$ as functions of $J$ with $\protect\rho =0.75$ and $U=50.0$%
.\ }
\label{fig2}
\end{figure}

\begin{figure}[b]
\includegraphics[width=9cm]{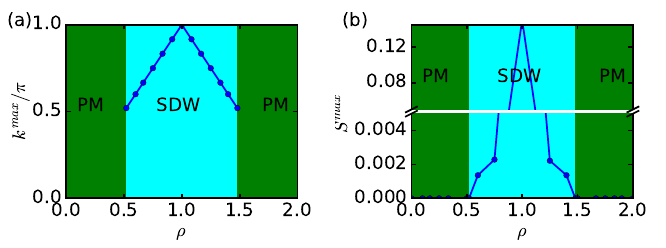}
\caption{(a) The ordering wave vector $k^{\max }$ and (b) the ordering
strength $S^{\max }$ as functions of $\protect\rho $ with $J=10.0$ and $%
U=50.0$.\ }
\label{fig3}
\end{figure}

This process can be verified by monitoring the ordering wave vector $k^{\max }$
and strength $S^{\max }$ as functions of $J$. Because of the inherent
inversion symmetry of $S(k)$ about $k=0$, we hereafter restrict our
discussion on the $k\geqslant 0$ part. As depicted in Fig.~\ref{fig2}%
, for relatively small coupling strength $J<7.0$, $S^{\max }$ is
extrapolated to zero and the corresponding $k^{\max }$ is not well defined,
signaling the existence of PM. A SDW phase occurs for $7.0<J<13.0$, in which both
$S^{\max }$ and$\ k^{\max }$ take nonzero values. Tuning the coupling
strength above $J=13.0$,\ $k^{\max }$ drops to zero, which is accompanied by
monotonically increased $S^{\max }$, implying the formation of FM. It should
be noticed that, within the SDW phase, $k^{\max }$ keeps at a constant value $%
0.75\pi $ irrespective of $J$. In fact, $k^{\max }$ in
this phase is uniquely set by the bosonic density. Figure \ref{fig3} demonstrates $k^{\max }$ and $S^{\max }$ in terms of $\rho $ for fixed $J$ and $U$, where we find that the relation
\begin{equation}
k^{\max }=|\pi \rho +2n\pi |  \label{km}
\end{equation}
is perfectly satisfied provided that the SDW phase is reached. Note that in Eq.~(\ref{km}),
the integer number $n$ ($=0,\pm 1,\pm 2...$) is introduced to place the
value of $k^{\max }$ into the first Brillion zone $-\pi \leqslant k\leqslant
\pi $.

These results encourages us to relate the bosonic SDW state with the Peierls state which
usually emerges in the 1D fermionic system with lattice degrees of freedom%
~\cite{Peierls1}. According to the Peierls theory in 1D, the ordering wave
vector in a Peierls state is connected to the Fermi wave vector $k_{F}$ via $%
k^{\max }=2k_{F}=\pi \rho $, due to the perfectly nested Fermi surfaces~\cite%
{Superradiance,Fradkin13}. Remarkably, this relation also holds true in the
SDW phase of our bosonic model, despite the absence of a Fermi surface. This suggests that a more comprehensive theoretical framework,
capable of unifying the fermionic and bosonic cases, may be necessary to
fully understand Peierls transitions.

\begin{figure}[t]
\includegraphics[width=9cm]{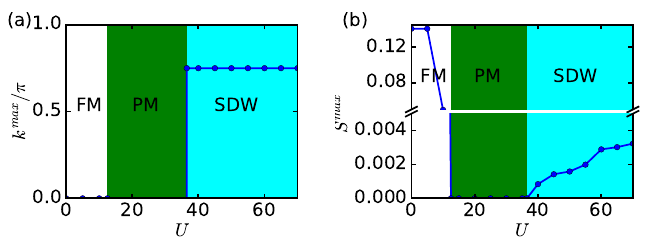}
\caption{(a) The ordering wave vector $k^{\max }$ and (b) the ordering
strength $S^{\max }$ as functions of $U$\ for systems with\ $J=10.0$ and $%
\protect\rho =0.75$.}
\label{fig4}
\end{figure}

The bosonic Peierls transition is not a perturbation effect, as can be
inferred in Sec. \ref{sec:perturbation theory}. Instead, it is essentially a
strongly correlated effect which depends on the competition between
relatively large $U$ and $J$. This is in contrast with the 1D fermionic
system, where the direct interactions between fermions is irrelevant to the
formation of a Peierls state~\cite{Fradkin13}. The effect of the bosonic
Hubbard interaction is illustrated in Fig.~\ref{fig4}, where we plot the
variation of $k^{\max }$ and $S^{\max }$ in terms of $U$ for $\rho =0.75$.
As can be seen, the SDW phase can be formed only when the interaction is larger
than some critical value $U_{\text{c}}\approx 37.0$. \ It is observed again
that the ordering wave vector of the SDW remains at $0.75\pi(=\pi \rho)$ irrespective of $U$, a principal feature of the Peierls state.

\begin{figure*}[tp]
\includegraphics[width=17.0cm]{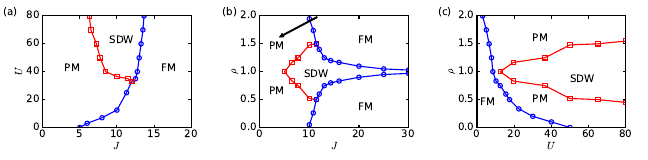}
\caption{(a) The phase diagram in the $J-U$ plane with $\protect\rho =0.75$,
(b) the phase diagram in the $J-\protect\rho $ plane with $U=50.0$, and (c)
the phase diagram in the $U-\protect\rho $ plane with $J=10.0$.\ The phase
boundaries in these phase diagrams have been extrapolated to the
thermodynamic limit $L\rightarrow \infty $.}
\label{fig5}
\end{figure*}

The competition of $U$ and $J$ for fixed density $\rho =0.75$\ is
quantitatively reflected in the $J-U$ phase diagram in Fig.~\ref{fig5}(a). We can
clearly find a lower bound of $U$, below which the system evolves directly
from PM to FM as $J$ increases. The transition from PM to FM becomes easier
for smaller $U$, consistent with the perturbation theory we developed in
Subsection \ref{sec:weak}. Above the lower bound of $U$, the bosonic Peierls
phase, which is more favored as $U$ increases, appears in between the PM and
FM.

Given that the nature of the SDW phase is intimately connected to the bosonic
density, it is then instructive to investigate the effect of density on various magnetic transitions. We map out the phase
diagram in the $J-\rho $ plane with $U=50$ in Fig.~\ref{fig5}(b). At any
fixed $\rho $, a general feature of the ground state is that it tend to be
PM and FM for two opposite limit $J\ll 1$ and $J\gg 1$, respectively. The
SDW order is stabilized at an intermediate value of $J$ instead, and it is more
favored when the system is closed to unit filling $\rho =1$. The formation
of FM at low bosonic densities can be understood by noting that the
conduction bosons are sparsely distributed on the lattice, in which case the
interaction $U$ is irrelevant. Because of a large antiferromagnetic Kondo
coupling, different spin configurations of the localized moments behave like
energy barriers hindering the motion of the conduction bosons. The most
energetically favorable choice is to ferromagnetically align the localized
spins, such that the bosons can freely move along the lattice, which lowers
the energy by $\Delta E\sim t$.

\begin{figure}[b]
\includegraphics[width=9.5cm]{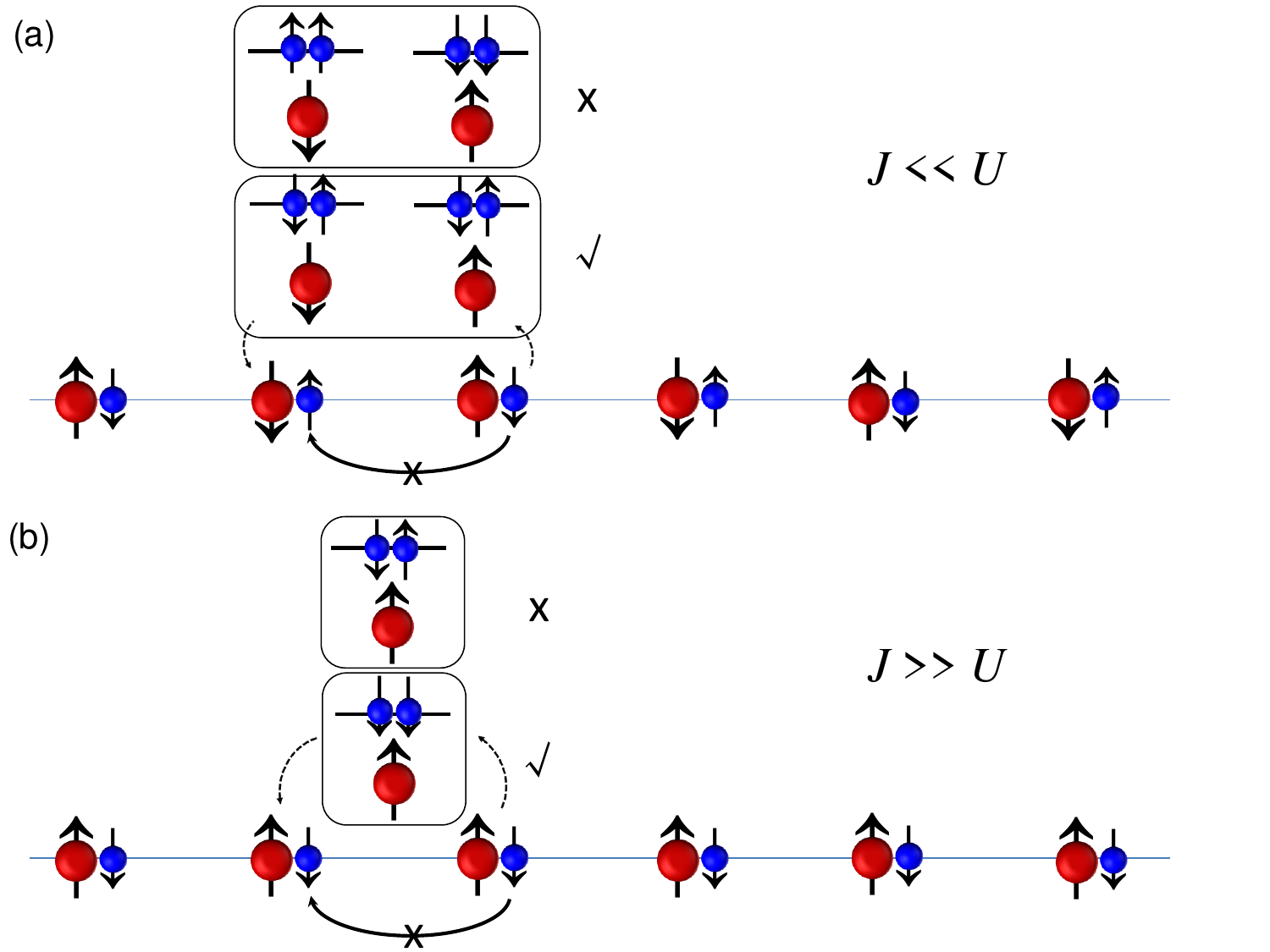}
\caption{Schematic illustration of the magnetic structures and the
corresponding superexchange mechanisms for (a) $J\ll U$ and (b) $J\gg U$. In these figures, the large red and small blue balls denote the localized moments and conduction electrons, respectively. The arrows in the balls indicate the spin orientations.}
\label{fig6}
\end{figure}

Nevertheless, the competition between $J$ and $U$ becomes crucial if the
system is tuned to be around unit filling.\ In this case, each neighbor
sites of the conduction bosons is likely to be occupied by another one. The
direct tunneling of the conduction bosons between different sites is prohibited
due to a large energy penalty costed by $J$ and $U$. However, the energy can
still be lowered, through exchanging excitations with some virtual states, by $%
\Delta E\sim t^{2}/J$ ($\Delta E\sim t^{2}/U$), once the conduction bosons
form an antiferromagnetic (ferromagnetic) spin configuration. From the
viewpoint of superexchange mechanism, a relatively large $U$ ($J$) is amount
to imposing a constraint on the intermediate virtual states such that double
occupancy of the conduction bosons with parallel (antiparallel) sspin is
excluded from the Hilbert space (see Fig.~\ref{fig6} for illustration). It
thus follows that the SDW phase (FM) is energetically more favorable for at least $%
J\lesssim U$ ($J\gtrsim U$). This feature becomes clearer in the $U-\rho $
phase diagram, as we show in Fig.~\ref{fig5}(c). It is seen that, while the
FM is generally stabilized at small $U/J$, the SDW phase can be formed only when $%
U\gtrsim J$ and the bosonic density is not far away from $\rho =1$. As $U$
increases, the range of $\rho $, where the SDW order can exist,\ grows up. Notice
that at unit filling, the critical point of $U$, above which the FM is destabilized, is estimated as $U_{\text{c}}\approx 7.5$, in agreement with
the perturbation results we derived in Subsection \ref{sec:strong}.

\begin{figure}[b]
\includegraphics[width=6.0cm]{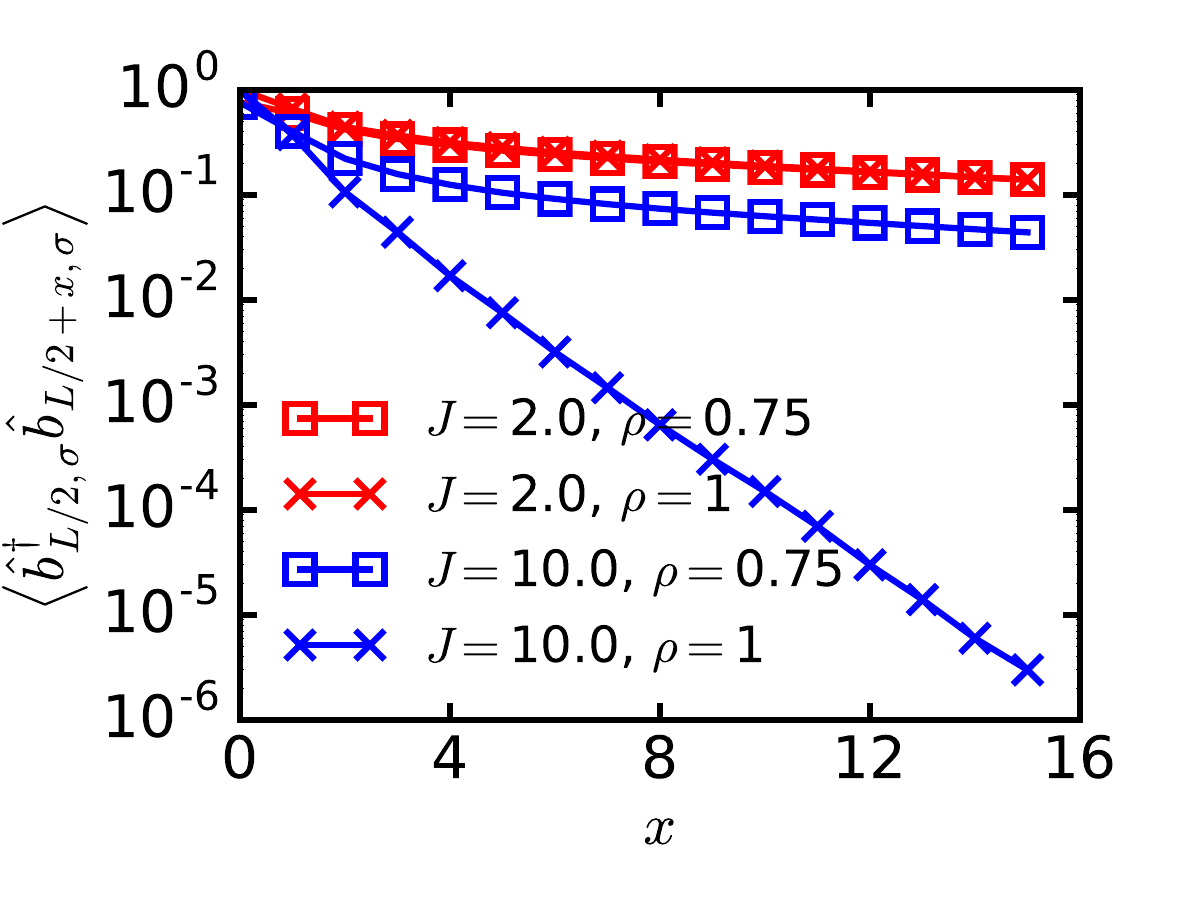}
\caption{The superfluid correlation $\left\langle \hat{b}_{L/2,\protect%
\sigma }^{\dag }\hat{b}_{L/2+x,\protect\sigma }\right\rangle $ as a function of the distance $x$ for different $J$ and $\rho$. The data for PM and SDW phase are labeled by red and blue curves,  respectively. The straight line for $J=10$ and $\rho=1$ shows that the decay in this case is exponential. The other parameters are $U=50.0$, $\protect\rho =0.75$, and $L=32$.}
\label{fig7}
\end{figure}

Another important feature accompanied with the SDW order is the absense of the off-diagonal bosonic long-range orders at commensurate fillings. The superfluid correlation
function $\left\langle \hat{b}_{L/2,\sigma }^{\dag }\hat{b}_{L/2+x,\sigma
}\right\rangle $ for different densities $\rho $\ and Kondo couplings $J$ are depicted in Fig.~\ref{fig7}. It is shown that the correlation $%
\left\langle \hat{b}_{L/2,\sigma }^{\dag }\hat{b}_{L/2+x,\sigma
}\right\rangle $ with $J=2.0$ and different $\rho$ presents long-range order, which, as expected,
inherits the superfluid nature of the Bose-Hubbard model. For larger Kondo couplings at which the SDW order is stabilized, on the other hand, the property of the superfluid correlation depends strongly on the commensurability of $\rho$.  As plotted by blue curves in  Fig.~\ref{fig7}, while the long-range feature of the superfluid correlation remains at incommensurate fillings (i.e., $\rho$ takes non-integer values), $\left\langle \hat{b}_{L/2,\sigma }^{\dag }\hat{b}_{L/2+x,\sigma
}\right\rangle $ at unit filling decreases exponentially with the distance $x$, indicating the absence of  the off-diagonal bosonic long-range orders.

\begin{figure}[t]
\includegraphics[width=9.0cm]{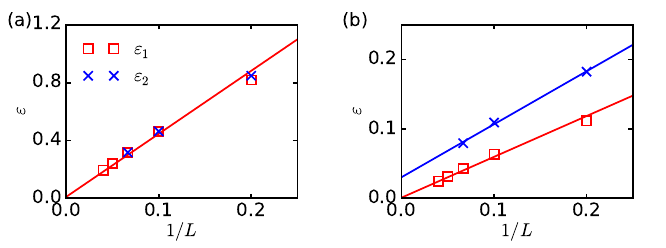}
\caption{The lowest excitation energy $\protect\varepsilon _{1}$ (red
square)\ and the second lowest excitation energy $\protect\varepsilon _{2}$
(blue cross) as a function of the inverse chain length in (a) the PM for $%
J=2.0$ and (b) the SDW phase for $J=10.0$.\ The other parameters are set by $U=80.0$
and $\protect\rho =0.8$.}
\label{fig8}
\end{figure}

\begin{figure}[b]
\includegraphics[width=9.0cm]{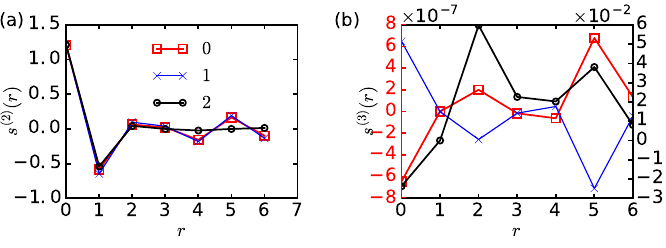}
\caption{(a) The two-body spin correlation $s^{(2)}(r)$ and (b) the
three-body spin correlation $s^{(3)}(r)$ with $J=10.0$, $U=80.0$, $\protect%
\rho =0.8$, and $L=15$.\ }
\label{fig9}
\end{figure}

It is known that a Peierls gap naturally emerges in the fermionic
density wave phase~\cite{Peierls1,Holstein}. This feature is also present in the bosonic Peierls state here.  To see this clearly, one can analyze the scaling of the lowest excitation energies $%
\varepsilon _{n}=E_{n}-E_{0}$ with the system size, where $E_{n}$ is the
energy of the $n$th lowest eigenstate of the Hamiltonian (\ref{H1}). In the
PM phase with superfluid character, we have found that the $\varepsilon _{n}$
decrease rapidly with the system size and vanishes in the thermodynamic
limit, as seen in Fig.~\ref{fig8}(a). This implies that the ground state
here is unique and gapless, as expected for a standard superfluid. However, the
scaling of $\varepsilon _{n}$ in the SDW phase exhibits completely different
behavior. As shown in Fig.~\ref{fig8}(b), while $\varepsilon _{1}$ is
extrapolated to zero as $L\rightarrow \infty $, the energy differences
between the two lowest eigenstates and the higher excited states remain
finite in the thermodynamic limit. This means that the ground state of the
SDW is twofold degenerate in the $L\rightarrow \infty $ limit. This
degeneracy originates from the $Z_{2}$ symmetry of the system with respect
to the spin flip transformations $\hat{s}_{j}^{z}\rightarrow -\hat{s}_{j}^{z}
$ and $\hat{S}_{j}^{z}\rightarrow -\hat{S}_{j}^{z}$. To demonstrate this, we
have calculated the two- and three-body spin correlations, $%
s^{(2)}(r)=\sum_{l}\left\langle \hat{s}_{l}^{z}\hat{s}_{l+r}^{z}\right\rangle $ and $s^{(3)}(r)=\sum_{l}\left\langle \hat{s}_{l}^{z}\hat{s}%
_{l+1}^{z}\hat{s}_{l+1+r}^{z}\right\rangle $, respectively, under the three
lowest eigenstates in the SDW phase. As can be seen in Figs.~\ref{fig9}(a) and \ref%
{fig9}(b), while $s^{(2)}(r)$ calculated for the first exited state tends to be
the same value as for the ground state, the quantity under the second exited state shows dramatically different configuration. Further more, the three-body correlation $s^{(3)}(r)$ for the two lowest eigenstates presents exactly inverted patterns, which is also quite different from that for the higher eigenstates. Therefore, both the two lowest eigentates are bosonic Peierls states with finite excitation gaps in the thermodynamic limit. It is thus interesting to find that, although possessing both diagonal and off-diagonal long-range correlations, the bosonic Peierls phase at incommensurate fillings is different from the standard supersolid because of its gapped nature~\cite{Hubbard}.


\begin{figure}[b]
\includegraphics[width=6.0cm]{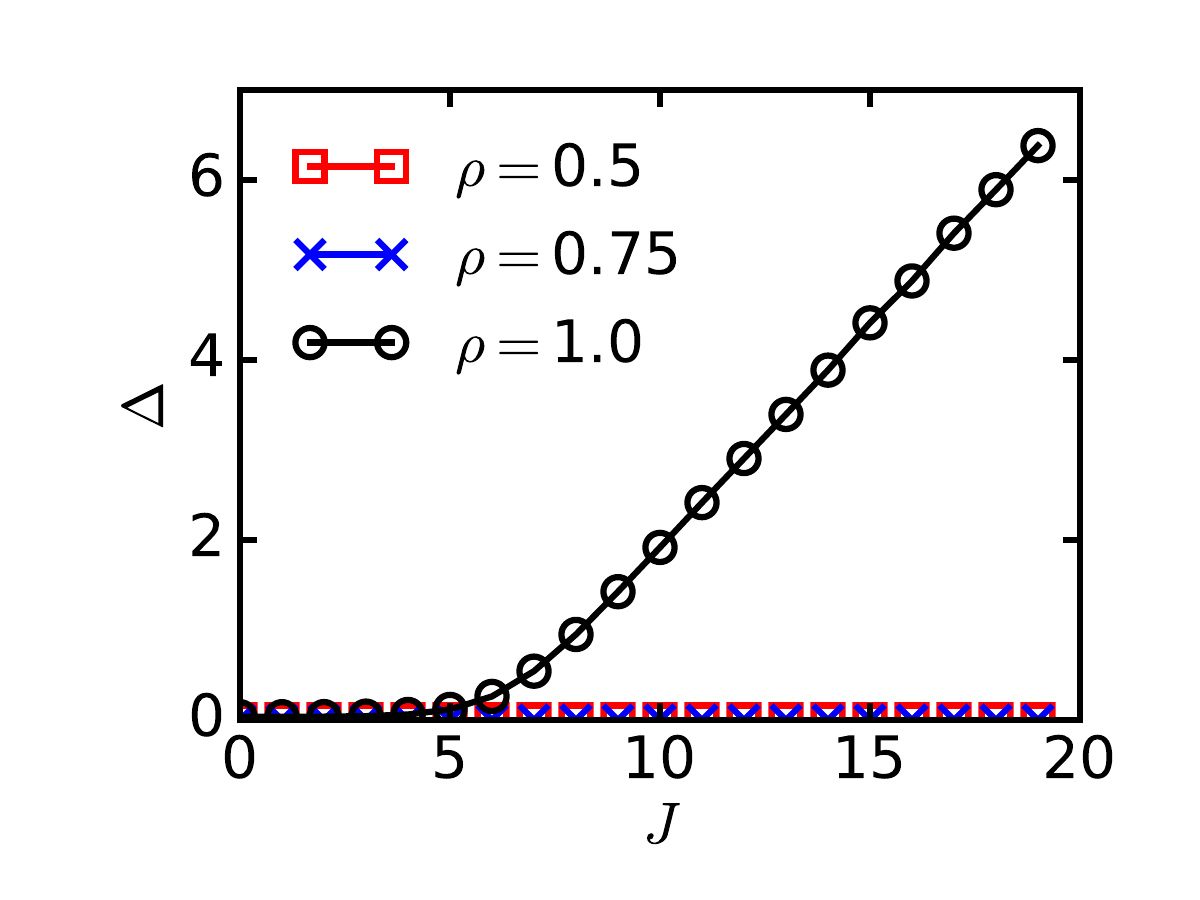}
\caption{The charge gap $\Delta $, extrapolated to $L\rightarrow \infty $,
as a function of $J$ for $U=80.0$ and different $\protect\rho $.}
\label{fig10}
\end{figure}

The appearance of diagonal density wave orders in the ground state is often
accompanied with a nonzero charge gap, i.e., the state is incompressible
with zero compressibility $\kappa =\partial \rho /\partial \mu =0$. Figure %
\ref{fig10} plots the charge gap $\Delta $, which has been extrapolated to
the $L\rightarrow \infty $ limit, in terms of $J$ for $U=50$ and different $\rho $. It is observed that at commensurate bosonic fillings, a
finite charge gap indeed exists in the SDW phase, consistent with the results
from perturbation theory (see Subsection \ref{sec:strong}). However, the SDW phase
with incommensurate bosonic fillings turns out to be compressible with
vanishingly small charge gap. Therefore, although the ground state keeps
gapped throughout the SDW phase, its compressibility depends crucially on the
bosonic density. The concurrence of a excitation gap, a diagonal long-range
order, and a nonzero compressibility can be understood under the
single-particle fermionic picture. When adding or removing particles from
the fermi-lattice system, the band structure, especially the position of
the gap, is modified at the same time according to the relation $k^{\max
}=2k_{F}$. This lattice relaxation effect can prevent the gap
penalty when changing the particle density, giving rise to a nonzero
compressibility. While the relation $k^{\max }=2k_{F}$ still holds
in the bosonic Peierls state, the underlying mechanism can not be understood
on a single-particle level, as the Fermi surface picture is lacking \cite{BP1}. Its
microscopic origin for general particle densities should be traced back to
the strong boson-boson and boson-lattice correlation, the exploration of
which is beyond the scope of this paper and we leave to future work.

\begin{figure}[b]
\includegraphics[width=7.0cm]{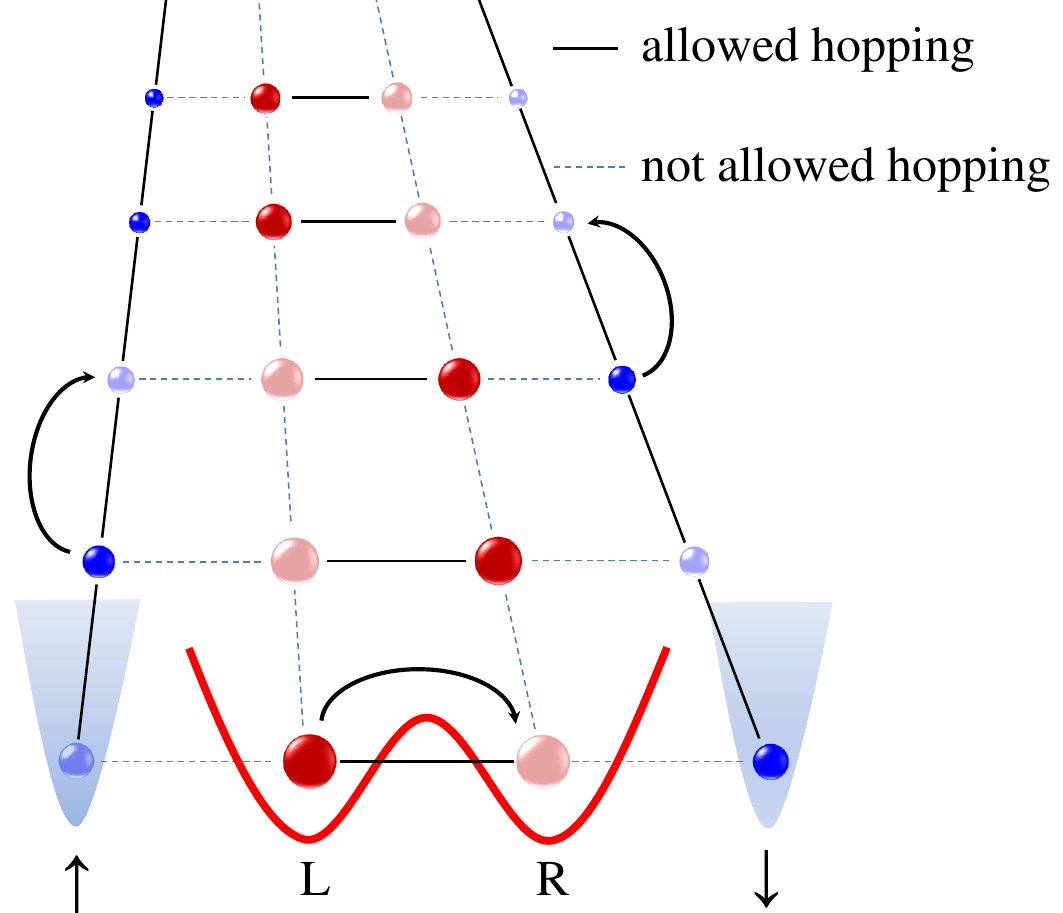}
\caption{Schematic representation of the experimental setup used to
implement the bosonic IKL Hamiltonian. A gas of ultracold bosonic atoms,
trapped by two 1D optical lattices, simulate the conduction bosons (the small blue balls). The
locations of the two 1D optical lattices can be used to represent the spin
index of the conduction bosons. The atoms can hop along each individual 1D
lattice, but the hopping between the two lattices is not allowed. The other
two atomic chains (the large red balls), placed in between the original two 1D optical lattices,
are tightly trapped such that the particle hoping along each chain is
forbidden but the interchain hopping is allowed. These localized atoms can
be utilized to simulate the localized magnetic moments in the IKL. In this figure, particle hoppings along the bonds labeled by solid (dashed) lines are allowed (not allowed).}
\label{fig11}
\end{figure}

\section{Possible experimental implementation}

The most convenient way to realize the bosonic IKL is to properly control
ultracold bosonic atoms in optical lattices. The conduction bosons and
localized magnetic moments can be realized by loading atoms in energy bands
with different mobilities~\cite{BHighOL1,BHighOL2,BHighOL3}, or populating
optical lattices by alkaline-earth-metal atoms which may experience
different trapping potentials, depending on their orbital angular momenta%
~\cite{HighOL3,HighOL4,HighOL5,HighOL6,HighOL7,HighOL8}. The Ising anisotropy
of the Kondo coupling can be easily achieved by various methods, including
the confinement-induced resonances~\cite{Anisotropy1} and the laser-induced
inter-channel coupling~\cite{Anisotropy2,Anisotropy3}. We here provide an
alternative way to experimentally implement our model. The basic idea is to
simulate the spin degrees of freedom by lattice sites in real space. As
shown in Fig.~\ref{fig11}, we consider\ first a gas of ultracold bosonic
atoms confined in two independent 1D chains, which are arranged to be
parallel to each other and can therefore represent different spin index of
the conduction bosons. Each chain of the bosonic atoms is described by the
Bose-Hubbard Hamiltonian. Such a scenario can be engineered by imposing an
optical superlattice~\cite{SuperOL1,SuperOL2}. Two additional chains of
ultracold atoms, tightly trapped either by optical tweezers or a second
optical superlattice with a double-well structure, are introduced in between
the original two potential minima of the conduction bosons. These tightly
trapped atoms, which we refer to as localized atoms, form a ladder structure
in real space. While the particle hopping along the two legs of the ladder
is forbidden by large energy barriers, a weak tunneling is allowed along the
rung. We further assume that each double well is occupied by only one
localized atom, which can be realized adiabatically in experiment. The
Hamiltonian describing the particle dynamics in such a configuration can be
readily written as

\begin{eqnarray}
H &=&-t\sum_{\left\langle i,j\right\rangle ,\sigma }\hat{b}_{i,\sigma
}^{\dag }\hat{b}_{j,\sigma }+\frac{U}{2}\sum_{j,\sigma }\hat{n}_{j,\sigma }(%
\hat{n}_{j,\sigma }-1)  \notag \\
&&+G\sum_{j}(\hat{n}_{j,\uparrow }\widehat{\bar{n}}_{j,L}+\hat{n}%
_{j,\downarrow }\widehat{\bar{n}}_{j,R})  \notag \\
&&+t_{\bot }\sum_{j}(\hat{c}_{j,L}^{\dag }\hat{c}_{j,R}+\hat{c}_{j,R}^{\dag }%
\hat{c}_{j,L})  \label{Hse}
\end{eqnarray}%
where $\widehat{\bar{n}}_{j,s}=\hat{c}_{j,s}^{\dag }\hat{c}_{j,s}$ with $%
\hat{c}_{j,s}^{\dag }$ ($\hat{c}_{j,s}$) the creation (annihilation)
operator of the localized atoms. The subscript $s$ ($=L,R$) labels the position of the localized atoms in each double well. The first two terms of the Hamiltonian (\ref%
{Hse}) are the Bose-Hubbard Hamiltonians for the conduction bosons, and the
third term describes the nearest neighbor interactions between the
conduction bosons and the\ localized atoms. The tunnelings of the localized
atoms in each double well are accounted for by the last term of the
Hamiltonian (\ref{Hse}). Making use of the schwinger transformations $\hat{s}%
_{j}^{z}=(\hat{n}_{j,\uparrow }-\hat{n}_{j,\downarrow })/2$, $\hat{S}%
_{j}^{z}=(\widehat{\bar{n}}_{j,L}-\widehat{\bar{n}}_{j,R})/2$, $\hat{S}%
_{j}^{x}=(\hat{c}_{j,L}^{\dag }\hat{c}_{j,R}+\hat{c}_{j,R}^{\dag }\hat{c}%
_{j,L})/2$, and the constraint $\widehat{\bar{n}}_{j,L}+\widehat{\bar{n}}%
_{j,R}=1$, the Hamiltonian (\ref{Hse}) immediately reduces to the bosonic
IKL Hamiltonian (\ref{H1}) with the substitution of the coupling constants $%
J\longrightarrow 2G$ and $h$ $\longrightarrow 2t_{\bot }$. In this sense,
the localized atoms play the role of localized magnetic moments in the bosonic IKL,
and the spin degrees of freedom are encoded in the two minima of each double
well. Hamiltonian (\ref{Hse}) has the advantage that all parameters can be
tuned independently. For example, $t$ and $t_{\bot }$ can be controlled by
the depths of the optical traps, and $U$ and $G$ can be tuned through the
Feshbach resonance~\cite{Feshbach} or the confinement induced resonance~\cite%
{Anisotropy1}.

\section{Conclusion}

In conclusion, we have studied a bosonic version of the 1D IKL, in which the
electrons are replaced by spin-$1/2$ bosons. Utilizing a perturbative
analysis and the numerical DMRG
method, we have characterized the phases of the system in different
parameters regimes. Apart from the PM and FM, a Peierls-like state,
characterized by a long-range SDW order and a nonzero excitation gap, has
been revealed. We have also discussed the possibility of implementing the
model using ultracold atoms in optical lattices.

\section*{Acknowledgments}

This work is supported by the National Key R\&D Program of China under Grant
No. 2022YFA1404201, the National Natural Science Foundation of China (NSFC)
under Grant No.~12174233, 12034012 and 12004230, the Research Project
Supported by Shanxi Scholarship Council of China and Shanxi '1331KSC'. Our
simulations make use of the ALPSCore library~\cite{ALPS}, based on the
original ALPS project~\cite{ALPS2}.

\vbox{\vskip1cm} \appendix

\section{Derivation of the effective spin Hamiltonian in the strong coupling
limit}

We here derive the effective Hamiltonian (\ref{H1}) in the main text based
on the second order perturbation theory. Given that the effective
Hamiltonian involves only on-site and nearest-neighbor interactions, our
derivation will start from a two-site system, the extension of which to a
lattice is straightforward. Since we consider the strong coupling limit with
$U\gg t$ and $J\gg t$, it is convenient to write the original two-site
Hamiltonian as the addition of two parts,%
\begin{equation*}
\hat{H}_{\text{12}}=\hat{H}_{\text{12}}^{(\text{0})}+\hat{V}_{\text{12}}%
\text{,}
\end{equation*}%
where
\begin{equation*}
\hat{H}_{\text{12}}^{(\text{0})}=\frac{U}{2}\sum_{j=1,2,\sigma =\uparrow
,\downarrow }\hat{n}_{j,\sigma }(\hat{n}_{j,\sigma }-1)+J\sum_{j=1,2}\hat{s}%
_{j}^{z}\hat{S}_{j}^{z}
\end{equation*}%
and%
\begin{equation*}
\hat{V}_{\text{12}}=-t\sum_{\sigma =\uparrow ,\downarrow }(\hat{b}_{1,\sigma
}^{\dag }\hat{b}_{2,\sigma }+\text{H.c.})+h\sum_{j=1,2}\hat{S}_{j}^{x}\text{.%
}
\end{equation*}%
Here, $\hat{H}_{\text{12}}^{(\text{0})}$ is the zeroth order Hamiltonian which can
be treated exactly and $V_{\text{12}}$ denotes the perturbation part. The
state basis spanning the whole Hilbert space can be chosen as the product
states $\left\vert \hat{n}_{1,\uparrow },\hat{n}_{1,\downarrow };\bar{\sigma}%
_{1}\right\rangle \otimes \left\vert \hat{n}_{2,\uparrow },\hat{n}_{2,\downarrow };\bar{%
\sigma}_{2}\right\rangle $. We further assume the commensurate filling $\rho
=\hat{n}_{j,\uparrow }+\hat{n}_{j,\downarrow }$ on each site, and then the
ground-state manifold is constituted by states $\left\vert G_{\bar{\sigma}%
_{1},\bar{\sigma}_{2}}\right\rangle =\left\vert \rho /2+\bar{\sigma}%
_{1}m,\rho /2-\bar{\sigma}_{1}m;-\bar{\sigma}_{1}\right\rangle \otimes
\left\vert \rho /2+\bar{\sigma}_{2}m,\rho /2-\bar{\sigma}_{2}m;-\bar{\sigma}%
_{2}\right\rangle $, where the magnetization $m$ is fixed by Eq. (\ref{Mz}).
Note that by this notation, the antiferromagnetic nature of the Kondo
coupling has been taken into consideration. The state $\left\vert G_{\bar{%
\sigma}_{1},\bar{\sigma}_{2}}\right\rangle $ is fourfold degenerate with
respect to the free choice of $\bar{\sigma}_{1}$ and $\bar{\sigma}_{2}$.
This degeneracy can be lifted by virtual transitions to some exited states $%
\left\vert EX\right\rangle $. These virtual exited states can either be
charge-like or spin-like, depending on the ways they are connected to the
ground states. The charge-like (spin-like) exited states can be
straightforwardly obtained by acting the hopping (transverse field) term of $%
V_{\text{12}}$ on the states $\left\vert G_{\bar{\sigma}_{1},\bar{\sigma}%
_{2}}\right\rangle $. Using this principle, we can write the charge-like
exited states and their corresponding energies $E_{\text{ex}}$ as

\begin{widetext}

\begin{equation}
\left\vert EX\right\rangle =\left\{
\begin{array}{c}
\left\vert \rho /2+\bar{\sigma}m,\rho /2-\bar{\sigma}m-1;-\bar{\sigma}%
\right\rangle \otimes \left\vert \rho /2-\bar{\sigma}m,\rho /2+\bar{\sigma}%
m+1;\bar{\sigma}\right\rangle , \\
\left\vert \rho /2+\bar{\sigma}m+1,\rho /2-\bar{\sigma}m;-\bar{\sigma}%
\right\rangle \otimes \left\vert \rho /2-\bar{\sigma}m-1,\rho /2+\bar{\sigma}%
m;\bar{\sigma}\right\rangle
\end{array}%
\right. \text{ }E_{\text{ex}}=(2m+1)U  \label{EX1}
\end{equation}%



\begin{equation}
\left\vert EX\right\rangle =\left\{
\begin{array}{c}
\left\vert \rho /2+\bar{\sigma}m-1,\rho /2-\bar{\sigma}m;-\bar{\sigma}%
\right\rangle \otimes \left\vert \rho /2-\bar{\sigma}m+1,\rho /2+\bar{\sigma}%
m;\bar{\sigma}\right\rangle , \\
\left\vert \rho /2+\bar{\sigma}m,\rho /2-\bar{\sigma}m+1;-\bar{\sigma}%
\right\rangle \otimes \left\vert \rho /2-\bar{\sigma}m,\rho /2+\bar{\sigma}%
m-1;\bar{\sigma}\right\rangle
\end{array}%
\right. \text{ }E_{\text{ex}}=(1-2m)U+\frac{J}{2}  \label{EX2}
\end{equation}%



\begin{equation}
\left\vert EX\right\rangle =\left\{
\begin{array}{c}
\left\vert \rho /2+\bar{\sigma}m,\rho /2-\bar{\sigma}m-1;-\bar{\sigma}%
\right\rangle \otimes \left\vert \rho /2+\bar{\sigma}m,\rho /2-\bar{\sigma}%
m+1;-\bar{\sigma}\right\rangle , \\
\left\vert \rho /2+\bar{\sigma}m-1,\rho /2-\bar{\sigma}m;-\bar{\sigma}%
\right\rangle \otimes \left\vert \rho /2+\bar{\sigma}m+1,\rho /2-\bar{\sigma}%
m;-\bar{\sigma}\right\rangle , \\
\left\vert \rho /2+\bar{\sigma}m,\rho /2-\bar{\sigma}m+1;-\bar{\sigma}%
\right\rangle \otimes \left\vert \rho /2+\bar{\sigma}m,\rho /2-\bar{\sigma}%
m-1;-\bar{\sigma}\right\rangle , \\
\left\vert \rho /2+\bar{\sigma}m+1,\rho /2-\bar{\sigma}m;-\bar{\sigma}%
\right\rangle \otimes \left\vert \rho /2+\bar{\sigma}m-1,\rho /2-\bar{\sigma}%
m;-\bar{\sigma}\right\rangle
\end{array}%
\right. \text{ }E_{\text{ex}}=U\   \label{EX3}
\end{equation}%


Similarly, the spin-like exited states and energies are obtained as


\begin{equation}
\left\vert EX\right\rangle =\left\{
\begin{array}{c}
\left\vert \rho /2+\bar{\sigma}m,\rho /2-\bar{\sigma}m;-\bar{\sigma}%
\right\rangle \otimes \left\vert \rho /2+\bar{\sigma}m,\rho /2-\bar{\sigma}m;%
\bar{\sigma}\right\rangle , \\
\left\vert \rho /2+\bar{\sigma}m,\rho /2-\bar{\sigma}m;\bar{\sigma}%
\right\rangle \otimes \left\vert \rho /2+\bar{\sigma}m,\rho /2-\bar{\sigma}%
m;-\bar{\sigma}\right\rangle , \\
\left\vert \rho /2+\bar{\sigma}m,\rho /2-\bar{\sigma}m;-\bar{\sigma}%
\right\rangle \otimes \left\vert \rho /2-\bar{\sigma}m,\rho /2+\bar{\sigma}%
m;-\bar{\sigma}\right\rangle , \\
\left\vert \rho /2-\bar{\sigma}m,\rho /2+\bar{\sigma}m;-\bar{\sigma}%
\right\rangle \otimes \left\vert \rho /2+\bar{\sigma}m,\rho /2-\bar{\sigma}%
m;-\bar{\sigma}\right\rangle
\end{array}%
\right. \text{ }E_{\text{ex}}=mJ  \label{EX4}
\end{equation}%

\end{widetext}

With the exited states (\ref{EX1})-(\ref{EX4}) and the low energy states $%
\left\vert G_{\bar{\sigma}_{1},\bar{\sigma}_{2}}\right\rangle $, the
effective Hamiltonian can be represented in terms of the matrix elements of
the perturbation $\hat{V}_{\text{12}}$ via the formula~\cite{Counterflow}
\begin{equation}
\left\langle G_{\bar{\sigma}_{1},\bar{\sigma}_{2}}\right\vert \hat{H}_{\text{eff}%
}\left\vert G_{\bar{\sigma}_{1},\bar{\sigma}_{2}}\right\rangle
=-\sum_{\gamma }\frac{\left\langle G_{\bar{\sigma}_{1},\bar{\sigma}%
_{2}}\right\vert \hat{V}_{\text{12}}\left\vert \gamma \right\rangle
\left\langle \gamma \right\vert \hat{V}_{\text{12}}\left\vert G_{\bar{\sigma}%
_{1},\bar{\sigma}_{2}}\right\rangle }{E_{\gamma }-E_{0}}  \label{Eeff}
\end{equation}%
where $\gamma $ labels the exited states $\left\vert EX\right\rangle $, and $%
E_{\gamma }$ and $E_{0}$ represent the exited- and ground-state
eigenenergies of $\hat{H}_{\text{12}}^{(\text{0})}$, respectively. Inserting
Eqs. (\ref{EX1})-(\ref{EX4}) into Eq. (\ref{Eeff}), we can write the
effective Hamiltonian in terms of the spin operators $\hat{s}_{j}^{z}$ and $%
\hat{S}_{j}^{z}$, i.e.,
\begin{eqnarray}
\hat{H}_{\text{eff}} &=&\mathcal{J}\hat{s}_{1}^{z}\hat{s}_{2}^{z}+4m^{2}\mathcal{J}%
\hat{S}_{1}^{z}\hat{S}_{2}^{z}-2m\mathcal{J}(\hat{s}_{1}^{z}\hat{S}_{2}^{z}+%
\hat{s}_{2}^{z}\hat{S}_{1}^{z})  \notag \\
&&+\mathcal{J}^{\prime }(\hat{s}_{1}^{z}\hat{S}_{1}^{z}+\hat{s}_{2}^{z}\hat{S%
}_{2}^{z})\text{,}  \label{Hs12}
\end{eqnarray}%
where the coupling constants $\mathcal{J}$ and $\mathcal{J}^{\prime }$ are
defined in Eqs. (\ref{J})-(\ref{Jp}). Extending the two-site Hamiltonian (%
\ref{Hs12}) to a lattice, we immediately obtain the effective Hamiltonian (%
\ref{Hs}) in the main text.

\end{document}